\pdfoutput=1
\documentclass[12pt]{iopart}

\expandafter\let\csname equation*\endcsname\relax

\expandafter\let\csname endequation*\endcsname\relax

\usepackage{amsmath}
\usepackage{iopams}
\usepackage{amsfonts}
\usepackage{amssymb}
\usepackage{graphicx}

\begin{document}
\title[Polarization-based branch selection of bound states in the continuum]{Polarization-based branch selection of bound states in the continuum in dielectric waveguide modes anti-crossed by a metal grating}

\author{R KIKKAWA, M NISHIDA and Y KADOYA}
\address{Graduate School of Advanced Sciences of Matter, Hiroshima University, Higashihiroshima 7398530, Japan}
\ead{R. K. (rkikkawa93@gmail.com) or M. N. (mnishida@hiroshima-u.ac.jp)}

\date{\today}

\begin{abstract}
We investigate bound states in the continuum (BICs) in a planar dielectric waveguide structure consisting of a gold grating on a dielectric layer with a back layer of metal. In this structure, Friedrich-Wintgen (FW) BICs caused by the destructive interference between the radiations from two waveguide modes appear near the anti-crossing point of the dispersion curves. In this study, it is revealed that the branch at which the BIC appears changes according to the polarization of incident radiation. Based on a temporal coupled mode theory, it is shown that the BIC branch is determined by the sign of the product of the coupling coefficients between the two waveguide modes and external radiation, which is consistent with FW theory. The signs of the coupling coefficients are estimated by the waveguide-mode decomposition of the numerically obtained electric fields and are confirmed to vary depending on the polarization.
\end{abstract}


\maketitle

\clearpage

|
\section{Introduction}

A bound state in the continuum (BIC) [1,2] is a non-radiative state that lies in the continuous spectral range spanned by the radiation modes. Unlike the quasi-bound states, the BIC, in principle, has an infinitely high quality factor. Although the BIC itself cannot be excited by the incident wave, the Q factor becomes very high near the BIC, which is expected in high-Q devices. The BIC was originally mentioned in Quantum Mechanics in 1929 by Von-Neumann and Wigner [3]. The phenomenon is commonly seen in open-wave systems such as acoustics [4-6], water waves [7,8], and optics [1,9,10]. In optics, the BIC is attracting  interest in the context of metasurfaces [11-13], photonic structures [14-18] and plasmonics [19]. It has been shown that quite high Q-values are indeed realized near the BIC [9,20-22], and various nano-photonic/plasmonic applications using BICs have been proposed, such as lasers [23,24], modulators [25], and filters [26], SHG or nonlinear optics [27,28] etc.

 Most of the BICs found so far are categorized into two types: the symmetry-protected BIC [9,20,29,30] and the Friedrich-Wintgen (FW) BIC [10,14,15,17,31,32]. The former BIC results from the incompatible symmetry of one of the localized modes with the external radiation field that usually appears at the center of the Brillouin zone (at $\Gamma$ point). The latter results from the destructive interference of radiations from the two modes that can appear at the off-$\Gamma$ point around the crossing or anti-crossing point of the dispersion curves of the two modes, depending on the strength of the coupling between them.

We focus on the FW-BIC [32] formed between the two localized modes that are strongly coupled by the near-field overlap with the structural perturbation that produces the coupling to external radiation. In this case, a mode anti-crossing occurs, and the BIC appears on one of the two split-branches. Such a type of BIC is reported in cylindrical resonator system [33-35] and the periodic structures such as photonic [36-39] or plasmonic structures [40]. As is pointed out by Friedrich and Wintgen [32], the regime where the Q factor is kept high around the BIC becomes larger as the near-field coupling becomes stronger. This regime is called the near BIC [40] or the supercavity regime [35] and is targeted in searching for extraordinarily high-Q modes. Therefore, a thorough understanding of the physics behind this type of BIC is quite important for the realization of practical high-Q devices.

The position at which the FW-BIC appears is extremely sensitive to, and dependent on, the structure and parameters of the system. In FW theory [32], it was shown that the shift of the BIC position from the anti-crossing point is determined by the strength of the couplings among the two closed and one open channel based on the Feshbach projection operator method [41]. Moreover, which of the two interfering resonances becomes a BIC was shown to depend on the sign of the product of the matrix elements coupling these channels. Indeed, BICs have been observed on different branches depending on the system, e.g., on the lower energy branch in a metal relief grating coupled to a dielectric slab waveguide [40], and on the higher energy branch in a low-contrast dielectric grating [36]. Therefore, it is important to understand how the BIC position changes depending on the structure and parameters of the system, in order to characterize the interfering resonances and to determine the optimum design for a high-Q device.

In this study, we demonstrate a phenomenon where the BIC-branch selection is inverted by the polarization of incident light in a dielectric-waveguide/metal-grating structure. If the conclusion of FW theory is applicable to this system, this phenomenon is caused by the sign change of the coupling coefficients among waveguide modes and external radiation. We carry out two actions to confirm this hypothesis. Initially, following temporal coupled mode theory (TCMT) [42,43], which describes general classical open-wave systems, we obtain the BIC condition for the system with two resonant modes and one radiative mode. We pay attention to the signs of the coupling coefficients, and check whether the system conforms to FW theory. Next, we estimate the signs of the coupling coefficients by the waveguide-mode decomposition of the electric fields inside the dielectric layer, which is obtained numerically via the use of a spatial coupled mode theory (SCMT) [44-46], and observe how the sign of the product of the coupling coefficients depends on the polarization of incident light. We note here that SCMT and TCMT are essentially different theories despite their similar names. SCMT is an electromagnetic theory for those systems with a metallic aperture array, wherein Maxwell equations are solved by a modal expansion of electromagnetic fields. On the other hand, TCMT is a more general theory. This method derives a set of equations of motion for simple oscillators whose energy can be transferred to the outside. Therefore, no specific electromagnetic formulation is employed in it.

\section{Device structure}

We consider a planar dielectric waveguide structure as shown in Fig.\ 1. It consists of a gold grating on a dielectric layer with a backing metal layer of infinite thickness. Therefore, any wave incident from the grating side is totally reflected except for some absorption. The reflectance as well as the fields inside the device were calculated by the SCMT, for the incident plane wave with S or P polarization, of which the electric or magnetic field has only y-component. The grating with a period of  $\mathit{\Lambda}=433$ nm is composed of a gold strip whose width $w_{\mathrm{m}}$ is 0.98$\mathit{\Lambda}$ for the P-wave and 0.8$\mathit{\Lambda}$ for the S-wave. These widths were chosen to facilitate clear observation of the BICs. Note that the grating is 0th order in the air for the wavelength range considered here. To obtain the permittivity of gold, a Drude-Lorentz model fitted to the Johnson-Christy data [47] was used. 

\section{Calculation Results} 

We display the dispersion of the 0th order reflectance on the gray scale map for the two polarizations in Fig.\ 2. In both diagrams, dark bands (low reflectance = high absorption) are clearly seen. The dashed lines indicate the dispersion of the empty lattice modes of the dielectric waveguide sandwiched by the two flat metal planes. We name the lowest order waveguide modes TM$_{0}$ and TE$_{0}$, and the second modes, TM$_{1}$ and TE$_{1}$ for the P- and S-polarizations, respectively. The TE$_{1}$ mode is cut off for $\lambda>1.25$ ${\bf \mu}$m. The dark bands correspond well to the empty lattice modes. Therefore, they are attributed to absorption caused by the metallic loss associated with the dielectric waveguide modes.

At the intersection of the two modes, in both polarizations, one can see an anti-crossing that is caused by the near-field coupling between the two modes induced by the grating. Moreover, near the anti-crossing, the dark band disappeared locally at the positions indicated by the arrows in both polarizations. The disappearance of the absorption corresponds to the BIC. Here a remarkable feature is that the BIC lies on the lower-frequency branch for the P-polarization, whereas it lies on the higher-frequency branch for the S-polarization. This feature is the subject of this study and will be discussed in detail in the next section. However, we first confirm that the disappearance of the absorption is indeed due to the formation of the BIC by evaluating the imaginary part of the eigenfrequency along the upper and lower dark bands for the P-wave and S-wave cases, respectively. In the calculation, the imaginary part of the metal permittivity was removed so that the imaginary part of the eigenfrequency corresponds to the external (radiation) loss. The results are shown in Fig.\ 3. In both polarizations, the imaginary parts drop to zero at the point where the absorption disappears $(k_{x}\sim0.2\ [\pi/\mathit{\Lambda}])$, demonstrating that the BICs are indeed formed at that location.

\section{Discussion}

\subsection{Theoretical analysis using TCMT}

TCMT describes the behavior of resonances in a system in the time domain. In the present case, the system involves two resonant modes with a near field coupling and a far (radiation) field coupling, the latter via the 0th order grating corresponding to an input/output port. The time evolution of the two mode amplitude  $\mathbf{a}^\mathrm{T}=(a_{1} , a_{2})$ can be described by an effective Hamiltonian $\tilde{\mathbf{H}}$  and input wave $s_{+}$ with a coupling coefficient vector $\mathbf{D}=(d_{1} , d_{2})$ as 
\begin{equation}
\frac{d}{dt}\mathbf{a}=-i\tilde{\mathbf{H}}\mathbf{a}+\mathbf{D}^\mathrm{T}s_{+}=-i\{\mathbf{\Omega}-i(\mathbf{\Gamma}_{i}+\mathbf{\Gamma}_{e})\}\mathbf{a}+\mathbf{D}^\mathrm{T}s_{+},
\end{equation}
where
\begin{eqnarray}
\mathbf{\Omega}=\left(\begin{array}{@{\,}cccc@{\,}}\omega_{1} & \alpha \\ \alpha & \omega_{2} \\\end{array}\right), \mathbf{\Gamma}_{e}=\left(\begin{array}{@{\,}cccc@{\,}}\gamma_{e1} & \gamma_{0} \\ \gamma_{0}^* & \gamma_{e2} \\\end{array}\right) , \mathbf{\Gamma}_{i}=\left(\begin{array}{@{\,}cccc@{\,}} \gamma_{i1} & 0 \\ 0 & \gamma_{i2} \\\end{array}\right)
\nonumber
\end{eqnarray}
with $\omega_{1,2}, \gamma_{i1,2}$ and $\gamma_{e1,2}$ denoting the eigenfrequency, the internal loss, and the external loss of mode 1 or 2, respectively, and $d_{1,2}$ denoting the coupling between the external radiation and modes 1 and 2 through the port. The off-diagonal terms $\alpha$ in $\mathbf{\Omega}$  and $\gamma_{0}$  in $\mathbf{\Gamma}_{e}$  represent the near- and far-field coupling, respectively. Here, $\alpha$ is set to be real assuming that the effect of material loss on $\alpha$ is negligible, and that the system without material loss has time-reversal symmetry [48]. 
Considering the principle of the conservation of energy, the outgoing wave $s_{-}$ can be written as
\begin{equation}
s_{-}=c_{a}s_{+}+\mathbf{D\mathbf{a}}
\end{equation}
with $c_a$ representing the direct scattering coefficient.

From time-reversal symmetry and the energy conservation principle, the following relations are derived:
\begin{equation}
\mathbf{D}^\dagger\mathbf{D}=2\mathbf{\Gamma}_{e},
\end{equation}
\begin{equation}
c_{a}\mathbf{D}^*=-\mathbf{D}.
\end{equation} 
Using these relations, $\mathbf{D}$ is rewritten as
\begin{equation}
\mathbf{D}=(|d_1|e^{i\varphi_{1}},  |d_2|e^{i\varphi_{2}})=e^{i\varphi_{d}}(\sqrt{2 \gamma_{e1}},  p\sqrt{2 \gamma_{e1}})
\end{equation}
where $\varphi_{d}$ is an arbitrary phase and $p$ is a parity ($\pm1$) that represents the phase difference between $d_{1}$ and $d_{2}$. Here $p=1$  for the in-phase ($|\varphi_{1}-\varphi_{2}|=0$) case, and $p=-1$ for the anti-phase ($|\varphi_{1}-\varphi_{2}|=\pi$) case. Hereafter, the phase $\varphi_{d}$ is set to zero. This is always possible by adjusting the reference position for the external radiation. In addition, the off-diagonal term of $\mathbf{\Gamma_{e}}$ is rewritten as [42]
\begin{equation}
\gamma_{0}=p\sqrt{\mathbf{\gamma}_{e1}\mathbf{\gamma}_{e2}}.
\end{equation}  

The reflection coefficients for the incident wave with angular-frequency $\omega$ can be derived in TCMT from Eqs.\ (1) and (2) as
\begin{align}
r(\omega)&=c_{a}\bigg[1+ \nonumber \\ & 2\frac{\gamma_{e1}\{i(\omega_{2}-\omega)+\gamma_{i2}\}+\gamma_{e2}\{i(\omega_{1}-\omega)+\gamma_{i1}\}-i2p\alpha\sqrt{\gamma_{e1}\gamma_{e2}}} { \{(\omega_{1}-\omega)-i(\gamma_{i1}+\gamma_{e1})\}\{(\omega_{2}-\omega)-i(\gamma_{i2}+\gamma_{e2})\}+(i\alpha+p\sqrt{\gamma_{e1}\gamma_{e2}})^2 } \bigg],
\end{align}  
as  described in the Appendix. The reflectance spectra calculated using Eq.\ (7) are shown in Figs.\ 4(c) and 4(d) for $p=1$ and $-1$, respectively, and $\alpha>0$  for both cases. The parameters used in the calculations, listed in the caption, were decided so as to fit the SCMT results redrawn in Figs.\ 4(a) and 4(b) from Figs.\ 2(a) and 2(b) with some magnification. As is clearly seen, the reflectance spectra were accurately reproduced by the TCMT calculation. Importantly, the positions of the BIC are correctly predicted.

The appearance of the BIC is analyzed as follows [36,39,48]. Omitting the internal loss $\mathbf{\Gamma_{i}}$ , the eigenvalues for $\tilde{\mathbf{H}}$ are determined by
\begin{equation}
\left|\tilde{\mathbf{H}}-\omega\mathbf{I}\right|=
\{(\omega_{1}-\omega)-i\gamma_{e1}\}\{(\omega_{2}-\omega)-i\gamma_{e2}\}+(i\alpha+p\sqrt{\gamma_{e1}\gamma_{e2}})^2=0,
\end{equation}  
where $\mathbf{I}$ is the identity matrix. If we express the two solutions of Eq.\ (8) as $\beta$ and $\chi$, the sum and the product of them yield
\begin{eqnarray}
\beta+\chi&=&\omega_{1}+\omega_{2}-i\gamma_{e1}-i\gamma_{e2},\\
\beta\chi&=&(\omega_{1}-i\gamma_{e1})(\omega_{2}-i\gamma_{e2})+(i\alpha+p\sqrt{\gamma_{e1}\gamma_{e2}})^2.
\end{eqnarray}  	 
When a BIC is realized, one of the solutions is purely real. In that case, the solutions can be expressed using real numbers $A$ and $B$ as
\begin{eqnarray}
\beta&=&A-i\gamma_{e1}-i\gamma_{e2},\\
\chi&=&B.
\end{eqnarray}  	  
By substituting Eqs.\ (11) and (12) to Eq.\ (10), the following relations are obtained for $A$ and $B$,
\begin{eqnarray}
A+B&=&\omega_{1}+\omega_{2},\\
AB&=&\omega_{1}\omega_{2}-\alpha^2.
\end{eqnarray}  	   
Therefore, $A$ and $B$ are the solutions of the equation
\begin{equation}
x^2-(\omega_{1}+\omega_{2})x+\omega_{1}\omega_{2}-\alpha^2=0.
\end{equation}  	   
In addition, by comparing the imaginary parts of both sides of Eq.\ (10), the expression for $B$ is obtained as
 \begin{equation}
B=\frac{ \omega_{1}\gamma_{e2}+\omega_{2}\gamma_{e1}-2p\alpha\sqrt{\gamma_{e1}\gamma_{e2}} }{\gamma_{e1}+\gamma_{e2}}.
\end{equation}  
The solutions of Eq.\ (15) are
\begin{equation}
x_{\pm}=\frac{1}{2}\{\omega_{1}+\omega_{2}\pm\sqrt{(\omega_{1}-\omega_{2})^2+4\alpha^2}\}.
\end{equation}  	  
Because $B$ must be one of the two solutions, $x_{\pm}$, the following condition must be satisfied for Eq.\ (16);
\begin{equation}
\pm\sqrt{(\omega_{1}-\omega_{2})^2+4\alpha^2}=-\frac{(\omega_{1}-\omega_{2})(\gamma_{e1}-\gamma_{e2})+4p\alpha\sqrt{\gamma_{e1}\gamma_{e2}}}{\gamma_{e1}+\gamma_{e2}}.
\end{equation}  	  
By squaring both sides of Eq.\ (18), the following conditions are obtained for the existence of the BIC;
\begin{equation}
p\alpha(\gamma_{e1}-\gamma_{e2})=\sqrt{\gamma_{e1}\gamma_{e2}}(\omega_{1}-\omega_{2}).
\end{equation}  	
Substituting Eq.\ (19) into Eq.\ (8) and then solving Eq.\ (8) with respect to $\omega$, we obtain the following solutions .
\begin{equation}
\omega=
\left\{
\begin{array}{l}
\frac{1}{2}(\omega_{1}+\omega_{2})+\frac{p\alpha}{2}(\sqrt{\frac{\gamma_{e1}}{\gamma_{e2}}}+\sqrt{\frac{\gamma_{e2}}{\gamma_{e1}}})-i(\gamma_{e1}+\gamma_{e2}), \\ 
\frac{1}{2}(\omega_{1}+\omega_{2})-\frac{p\alpha}{2}(\sqrt{\frac{\gamma_{e1}}{\gamma_{e2}}}+\sqrt{\frac{\gamma_{e2}}{\gamma_{e1}}}).
\end{array}
\right.
\end{equation}  	
The latter solution represents the BIC, which does not have an imaginary part. Therefore, the BIC appears in the lower frequency branch when 
\begin{equation}
p\alpha>0,
\end{equation} 
whereas the BIC appears in the higher frequency branch when
 \begin{equation}
p\alpha<0.
\end{equation} 
This result is consistent with the FW theory [32].

The  regions where the BIC appears in the dispersion diagram can be discussed based on Eqs.\ (19), (21), and (22).
Consider the dispersion of $\omega_{1}$ and $\omega_{2}$ having positive and negative slopes, respectively as shown in Fig.\ 5. Let us first consider the case of $p\alpha>0$, for which, from Eq.\ (19), the signs of  $\omega_{1}-\omega_{2}$ and $\gamma_{1}-\gamma_{2}$ should be the same.
Hence, $\omega_{1}>\omega_{2}$ must be satisfied for $\gamma_{e1}>\gamma_{e2}$, corresponding to division (1) or (4) in Fig.\ 5(a), while $\omega_{1}<\omega_{2}$ for $\gamma_{e1}<\gamma_{e2}$, corresponding to division (2) or (3) in Fig.\ 5(b).
On the other hand, considering Eq.\ (21), the BIC appears on the lower branch for $p\alpha>0$.
Therefore, the BIC appears in division (4) in Fig.\ 5(a) for
$\gamma_{e1}>\gamma_{e2}$  and division (3) in Fig.\ 5(b) for $\gamma_{1}<\gamma_{2}$, respectively.
The case of $p\alpha<0$ can be analyzed in a similar way.
For $\gamma_{e1}>\gamma_{e2}$ and $\gamma_{e1}<\gamma_{e2}$, the BIC
appears in division (2) in Fig.\ 5(a) and division (1) in Fig.\ 5(b), respectively.
We further note that in both cases, the BIC appears on the mode with the lower radiative loss.

Next, let us check the correspondence between the above argument and the result in our structure shown in Fig.\ 4.
In the case of P-wave excitation shown in Fig.\ 4(a), we can see that the TM$_0$ ($\omega_{1}$) mode with positive slope is narrower than the TM$_1$ ($\omega_{2}$) mode with negative slope, namely $\gamma_{e1}<\gamma_{e2}$ corresponding to Fig.\ 5(b), and the BIC is located on the branch in division (3), as confirmed by the fitting shown in Fig.\ 4(c).
Hence, the above argument predicts $p\alpha>0$.
In the case of S-wave excitation, it is clear from Fig.\ 4(b) that the linewidth of the TE$_0$ ($\omega_{1}$) mode is much narrower than the TE$_1$ ($\omega_{2}$) mode, namely $\gamma_{e1}<\gamma_{e2}$ corresponding again to Fig.\ 5(b), and the BIC appears in division (1).
Therefore, $p\alpha<0$ is predicted.
In the next subsection, we will confirm the above prediction, $p\alpha>0$ and $p\alpha<0$ for the P-wave and S-wave excitations, respectively, in our specific structure, by checking the signs of $p$ and $\alpha$ from the SCMT calculation.

\subsection{Evaluation of the signs of $p$ and $\alpha$ using SCMT}

As  shown in the preceding section, the branch on which the BIC appears depends on the sign of $p\alpha$.
Here, $p$ denotes the phase difference (sign) between the coefficients that couple the waveguide modes with the external radiation; $p=1$ and $-1$ means $\varphi_{1}-\varphi_{2}=0$ and $\pi$, respectively.
The phase difference between the two waveguide modes excited by the incident radiation is determined solely by these coupling coefficients, if the effects of near- and far-field couplings are negligible.
Thus, the sign $p$ can be evaluated easily from the phases of the excited waveguide modes.
On the other hand, $\alpha$ represents the near field (direct) coupling between two resonant modes.
As shown below, the sign($\alpha$) can also be found by inspecting the phases of the waveguide modes that construct the coupled resonant modes at the anti-crossing point.

We determine the phase of the waveguide modes from the actual SCMT simulation. Assume that the electric and magnetic fields inside the dielectric layer is expanded by the propagating waveguide modes in the flat metal/dielectric/metal waveguide as [49]
\begin{eqnarray}
\mathbf{E}(x,z)&=&a_{1+}\mathbf{E}_{1+}(x,z)+a_{2+}\mathbf{E}_{2+}(x,z)+\cdots \nonumber \\ && + a_{1-}\mathbf{E}_{1-}(x,z)+a_{2-}\mathbf{E}_{2-}(x,z)+\cdots,\nonumber\\
\mathbf{H}(x,z)&=&a_{1+}\mathbf{H}_{1+}(x,z)+a_{2+}\mathbf{H}_{2+}(x,z)+\cdots \nonumber \\ && + a_{1-}\mathbf{H}_{1-}(x,z)+a_{2-}\mathbf{H}_{2-}(x,z)+\cdots.
\end{eqnarray}
Here, $\mathbf{E}_{i\pm}$, $\mathbf{H}_{i\pm}$ represents the transverse components (in $y$-$z$ plane in Fig.\ 1) of the electric and magnetic fields of the $i$-th waveguide mode, respectively, and the signs  $+$ and $-$ denote the mode propagating in the $+x$ and $-x$ directions, respectively.
$a_{i\pm}$ represents the complex amplitude of each waveguide mode.
As seen in Fig.\ 2, only two modes are relevant near the BIC point. Hence, Eq.\ (23) can be simplified to 
\begin{eqnarray}
\mathbf{E}(x,z)&=&a_{1s}\mathbf{E}_{1s}(x,z) + a_{2s'}\mathbf{E}_{2s'}(x,z), \nonumber\\
\mathbf{H}(x,z)&=&a_{1s}\mathbf{H}_{1s}(x,z) + a_{2s'}\mathbf{H}_{2s'}(x,z),
\end{eqnarray} 	  
where $s$ and $s'$ are either $+$ or $-$.
In the case of Fig.\ 2(a) for the P-wave radiation, mode 1 and mode 2 correspond to the TM$_0$ mode propagating in $+x$ direction and the TM$_1$ mode propagating $-x$ direction, respectively. In the case of Fig.\ 2(b) for the S-wave radiation, they correspond to  the TE$_0$ mode propagating in $+x$ direction and the TE$_1$ mode propagating in $+x$ direction, respectively.
The direction of propagation was determined from the slope of the corresponding empty lattice mode.
The amplitude $a_{is}$ is obtained by 
\begin{equation}
a_{is}=\int dz\left(\frac{\mathbf{E}\times\mathbf{H}_{is}^*+\mathbf{E}_{is}^*\times\mathbf{H}}{\mathbf{E}_{is}\times\mathbf{H}_{is}^*+\mathbf{E}_{is}^*\times\mathbf{H}_{is}}\right),
\end{equation}
using the electric field $\mathbf{E}$ and the magnetic field $\mathbf{H}$ obtained by the actual SCMT calculation. The integration is performed over the dielectric region.

\subsection{Evaluation of $p$}

We  evaluate the phase difference at the wavelength where the anti-crossing occurs, $k_{x}=0.1905\ [\pi/\mathit{\Lambda}]$ and $0.1600\ [\pi/\mathit{\Lambda}]$ for the P-wave and S-wave excitations, respectively.
The calculated phase difference in SCMT is shown in Figs.\ 6(a) and 6(b) for the P-wave and S-wave cases, respectively.
We denote the upper and lower branch of the coupled resonant wavelength as $\omega_{+}$ and $\omega_{-}$, respectively.
Around $\omega_{+}$ or $\omega_{-}$, the phase difference between the waveguide modes varies rapidly due to the influence of the excitation of each resonator.
However, in the region far from $\omega_{+}$ and $\omega_{-}$, the phase difference shows a convergence to 0 for the P-polarization and $\pi$ for the S-polarization.
In this region, the phase difference (sign) between the resonances should coincide with $p$, because  the effect of the resonances on the phase difference is considered to be negligible  (see Appendix ).

For comparison, the phase differences were also calculated in TCMT. Figure 6(c) is the phase difference for $p=1$ using the same parameters as for Fig.\ 4(c), and Fig.\ 6(d) is that for $p=-1$ with the parameters for Fig.\ 4(d).
The overall behavior of the phase difference in the SCMT results (Figs.\
6 (a) and (b)) corresponds very well to that in the TCMT results (Fig.\ 6 (c) and (d)).
Therefore, it can be concluded that $p=1$ and $p=-1$ for the P- and S-polarizations, respectively.

\subsection{ Evaluation of $\alpha$}

The sign of $\alpha$ can be deduced from the phase of the eigenmode (quasi-bound mode).
In our strong near-field coupling system, the eigenmode is mainly determined by $\alpha$ near the anti-crossing point.
In TCMT, by equating the eigenfrequencies of the two resonant modes, $\omega_{1}=\omega_{2}=\omega_{12}$, and neglecting $\mathbf{\Gamma_{i}}$ and $\mathbf{\Gamma_{e}}$, we have the eigenfrequencies $\omega_{+}$ and $\omega_{-}$ from
\begin{equation}
\left|\mathbf{\Omega}-\omega\mathbf{I}\right|=0,
\end{equation}
as
\begin{equation}
\omega_{\pm}=\omega_{12}\pm|\alpha|,
\end{equation}
and the eigenvectors are derived from
\begin{equation}
\left(\begin{array}{@{\,}cccc@{\,}}\omega_{12}-\omega & \alpha \\ \alpha & \omega_{12}-\omega \\\end{array}\right) 
\left(\begin{array}{c} a_{1}  \\ a_{2}  \\ \end{array} \right)=0.
\end{equation}
Hence, for $\alpha>0$, 
\begin{equation}
a_{1}=a_{2} \mbox{\ \ for\ \ } \omega_{+}, \quad a_{1}=-a_{2} \mbox{\ \ for\ \ } \omega_{-}
\end{equation}
and when $\alpha<0$ we have, 
\begin{equation}
a_{1}=-a_{2} \mbox{\ \ for\ \ } \omega_{+}, \quad a_{1}=a_{2} \mbox{\ \ for\ \ } \omega_{-}.
\end{equation} 
Therefore, the sign of $\alpha$ can also be evaluated from the phase difference arg($a_{1}$)-arg($a_{2}$) at $\omega_{+}$ or $\omega_{-}$ around the anti-crossing point. Here $\omega_{1}$ and $\omega_{2}$ correspond to either TM$_{0}$ (TE$_{0}$) or TM$_{1}$ (TE$_{1}$) for the P (S)-wave case.
We evaluated $a_{1}$ and $a_{2}$ from the electric field in the same way as 4.2.1 using Eq.\ (25). The electric field of the coupled resonant modes can be obtained numerically by searching for the electromagnetic wave solution in the absence of the incident wave using SCMT. In the calculation, we eliminated the metal loss of the whole structure.
Table 1(a) shows the results for the P-wave excitation at $k_{x}=0.1905$ $[\pi/\mathit{\Lambda}]$ and Table 1(b) shows those for the S-wave case at $k_{x}=0.1600$ $[\pi/\mathit{\Lambda}]$.
In both cases, the phase difference is about 0 at $\omega_{+}$ and $\pi$ at $\omega_{-}$, respectively.
Therefore, it can be concluded that $\alpha$ is positive for both polarizations .

\subsection{Sign of $p\alpha$}

As shown above, $p=1$ and $p=-1$ for the P- and S-polarizations, respectively and $\alpha$ is positive for both cases.
Although sign($p$) can be changed by the redefinition of the phase of one of the waveguide modes, this change must be accompanied by the change of sign($\alpha$) with sign($p\alpha$) unchanged [48].
Hence, we can conclude that sign($p\alpha$) is positive in the P-polarization and negative in the S-polarization in our structure, demonstrating that the prediction by the TCMT stated in Section 4.1 is correct.

\section{Conclusion}

In conclusion, we discussed, for the first time, the branch on which the FW-BIC appears in the anti-crossing dispersion of a photonic system with a simple and practically important structure, namely dielectric waveguide with metal grating. We demonstrated that the branch is selected by the incident polarization. The mechanism was explained by TCMT in terms of the polarization-dependent phase relation between the relevant waveguide modes forming the BIC. The polarization dependence of the BIC formation in our simple structure implies the external controllability of BICs in various optical and photonic devices. However, a question still remains whether the plasmonic nature plays an important role in our structure, which is important not only from a viewpoint of optical physics, but also for various applications, because non-plasmonic structures may be better to obtain higher Q values. Although an answer has not been obtained, our discussion based on the TCMT appplies regardless of the presence of the plasmonic effect, implying that the same or similar control of BICs is possible in various devices including all dielectric ones.

\section*{Acknowledgements}
This work was supported by JSPS KAKENHI Grant Numbers JP18K04979 and JP18K04980.

\appendix
\setcounter{section}{1}

\section*{Appendix}
Consider a case that $\mathbf{a}$ is time harmonic, $\mathbf{a}\propto e^{-i\omega t}$. Eq.\ (1) is written as 
\begin{equation}
[i(\mathbf{\Omega}-\omega\mathbf{I})+\mathbf{\Gamma}_{i}+\mathbf{\Gamma}_{e})]\mathbf{a}=\mathbf{D}^T s_{+}.
\end{equation}
Taking the inverse matrix from the left side of Eq.\ (A.1),
\begin{equation}
\mathbf{a}=[i(\mathbf{\Omega}-\omega\mathbf{I})+\mathbf{\Gamma}_{i}+\mathbf{\Gamma}_{e})]^{-1}\mathbf{D}^\mathrm{T}s_{+}.
\end{equation}
Substitute this into Eq.\ (2),
\begin{equation}
r(\omega) \equiv \frac{s_{-}}{s_{+}}=c_{a}+\mathbf{D}[i(\mathbf{\Omega}-\omega\mathbf{I})+\mathbf{\Gamma}_{i}+\mathbf{\Gamma}_{e})]^{-1}\mathbf{D}^\mathrm{T}.
\end{equation}
Under the condition of Eq.\ (3) and Eq.\ (4), inverse matrix is
\begin{align}
&[i(\mathbf{\Omega}-\omega\mathbf{I})+\mathbf{\Gamma}_{i}+\mathbf{\Gamma}_{e})]^{-1}= \nonumber \\ &\frac{ \left(\begin{array}{@{\,}cccc@{\,}}i(\omega_{1}-\omega)+\gamma_{i1}+\gamma_{e1} & i\alpha+p\sqrt{\gamma_{e1}\gamma_{e2}} \\ i\alpha+p\sqrt{\gamma_{e1}\gamma_{e2}} & i(\omega_{2}-\omega)+\gamma_{i2}+\gamma_{e2} \\\end{array}\right) } { \{(\omega_{1}-\omega)-i(\gamma_{i1}+\gamma_{e1})\}\{(\omega_{2}-\omega)-i(\gamma_{i2}+\gamma_{e2})\}+(i\alpha+p\sqrt{\gamma_{e1}\gamma_{e2}})^2 }. 
\end{align}  
After the matrix calculation on the RHS of Eq.\ (A.3), we finally get $r(\omega)$ as
\begin{align}
r(\omega)&=c_{a}\bigg[1+ \nonumber \\ & 2\frac{\gamma_{e1}\{i(\omega_{2}-\omega)+\gamma_{i2}\}+\gamma_{e2}\{i(\omega_{1}-\omega)+\gamma_{i1}\}-i2p\alpha\sqrt{\gamma_{e1}\gamma_{e2}}} { \{(\omega_{1}-\omega)-i(\gamma_{i1}+\gamma_{e1})\}\{(\omega_{2}-\omega)-i(\gamma_{i2}+\gamma_{e2})\}+(i\alpha+p\sqrt{\gamma_{e1}\gamma_{e2}})^2 } \bigg]. 
\end{align}  
On the other hand, the amplitudes of each mode can be obtained from Eq.\ (A.2) as
\begin{equation}
a_{1}=\frac{\{i(\omega-\omega_{2})-(\gamma_{i2}+\gamma_{i2})\}\sqrt{2\gamma_{e1}}+(p\sqrt{\gamma_{e1}\gamma_{e2}}+i\alpha)p\sqrt{2\gamma_{e2}}} { \{\omega-\omega_{1}+i(\gamma_{i1}+\gamma_{e1})\}\{\omega-\omega_{2}+i(\gamma_{i2}+\gamma_{e2})\}-(ip\sqrt{\gamma_{e1}\gamma_{e2}}-\alpha)^2 }s_{+} , \\
\end{equation}  
\begin{equation}
a_{2}=\frac{\{i(\omega-\omega_{1})-(\gamma_{i1}+\gamma_{i1})\}p\sqrt{2\gamma_{e2}}+(p\sqrt{\gamma_{e1}\gamma_{e2}}+i\alpha)\sqrt{2\gamma_{e1}}} { \{\omega-\omega_{1}+i(\gamma_{i1}+\gamma_{e1})\}\{\omega-\omega_{2}+i(\gamma_{i2}+\gamma_{e2})\}-(ip\sqrt{\gamma_{e1}\gamma_{e2}}-\alpha)^2 }s_{+} . \\
\end{equation}  
If $|\omega-\omega_{1}|$ and $|\omega-\omega_{2}|$ are much larger than the loss and the coupling coefficients, the amplitudes are approximated by 
\begin{equation}
a_{1}\simeq \frac{i\sqrt{2\gamma_{e1}}}{\omega-\omega_{1}}s_{+}, \\
\end{equation}  
 \begin{equation}
a_{2}\simeq p\frac{i\sqrt{2\gamma_{e2}}}{\omega-\omega_{2}}s_{+}. \\
\end{equation}  
Thus, the phase difference between $a_{1}$ and $a_{2}$ is determined by $p$ in the region far from resonances where the signs of $\omega-\omega_{1}$ and $\omega-\omega_{2}$ are the same.

\newpage

\clearpage

\noappendix

\begin{table}
\caption{Phase difference between the two waveguide modes to determine the sign($\alpha$). }
\vspace{3mm}
\begin{indented}
\item{(a)  P-wave at  $k_{x}=0.1905$ $[\pi/\mathit{\Lambda}]$.}
\item[]\begin{tabular}{@{}ll}
\br
Branch & $\mathrm{arg}(a_{\mathrm{TM1-}})-\mathrm{arg}(a_{\mathrm{TM0+}})$ \\
\mr
$\omega_{+}$ & $3.45\times10^{-4}\pi$ \\
$\omega_{-}$ & $0.997\pi$ \\
\br
\end{tabular}
\end{indented} 
\vspace{3mm}
\begin{indented}
\item{(b) S-wave at $k_{x}=0.1600$ $[\pi/\mathit{\Lambda}]$.}

\item[]\begin{tabular}{@{}ll}
\br
Branch & $\mathrm{arg}(a_{\mathrm{TE0+}})-\mathrm{arg}(a_{\mathrm{TE1+}})$ \\
\mr
$\omega_{+}$ &$-1.93\times10^{-2}\pi$ \\
$\omega_{-}$ &$0.989\pi$ \\
\br
\end{tabular}
\end{indented}
\end{table}

\clearpage

\begin{figure}
\centering
\includegraphics[width=7.5cm]{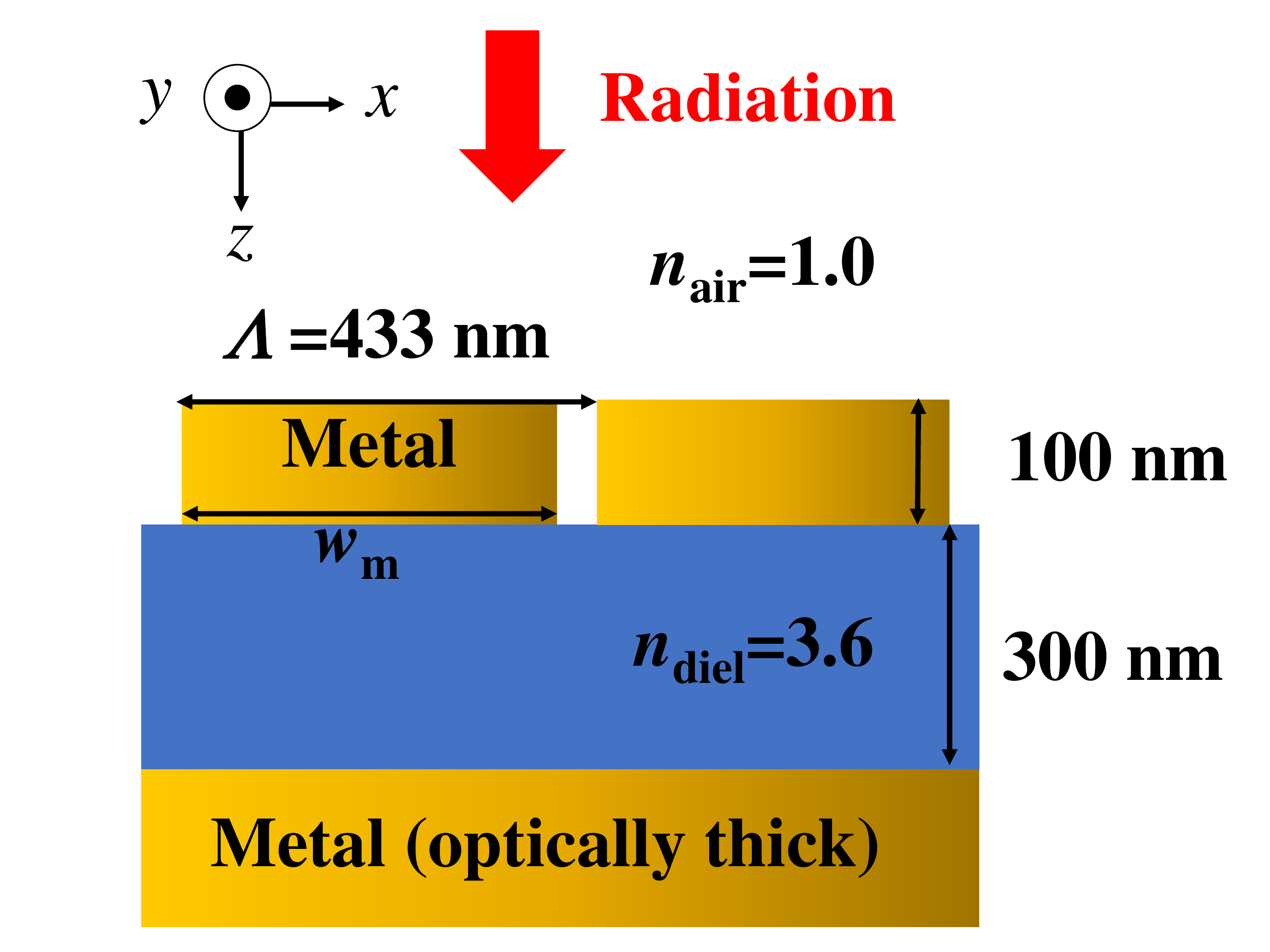}
\caption{Cross sectional structure of the device considered in this work.}
\label{fig:1}
\end{figure}

\begin{figure}
\centering
\includegraphics[width=15cm]{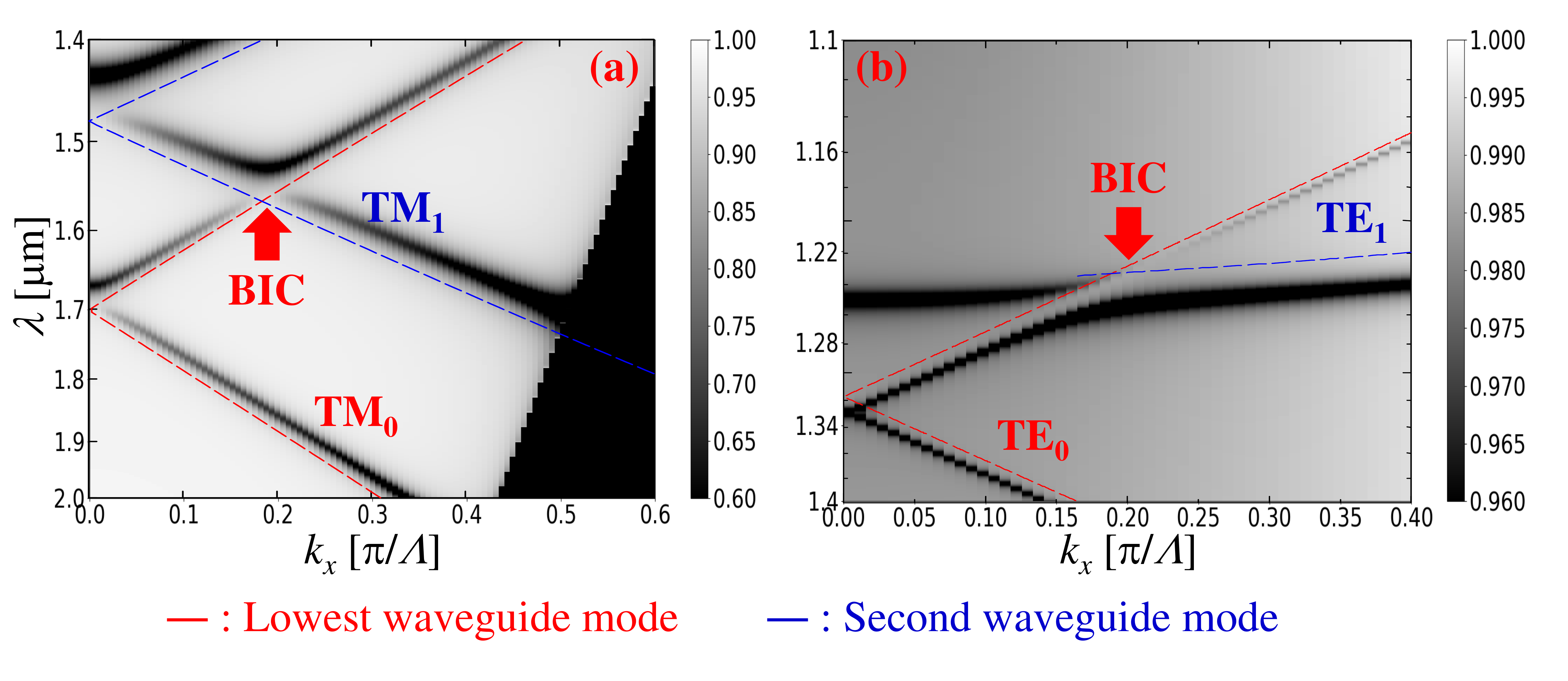}
\caption{ Reflectance  as a function of the in-plane wave vector and the incident wavelength; (a) P-wave incidence, (b) S-wave incidence. The red arrows indicate the location of the BICs.}
\label{fig:2a_2b}
\end{figure}

\begin{figure}
\centering
\includegraphics[width=15cm]{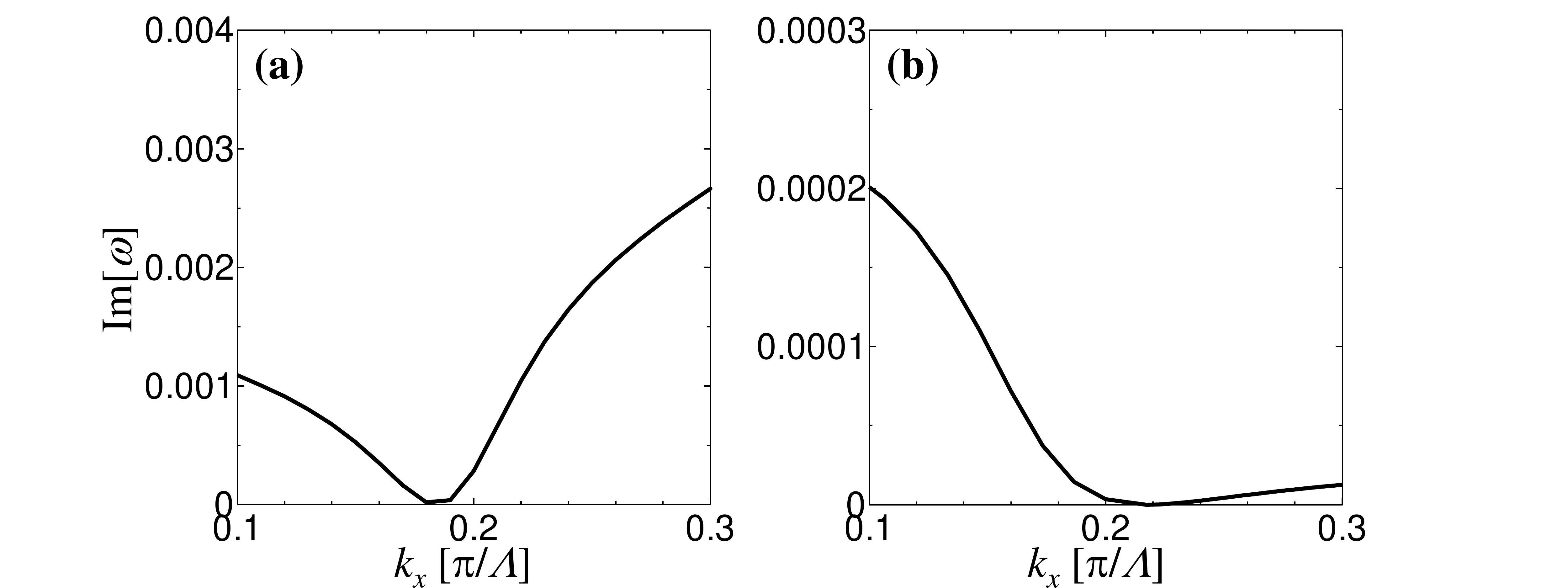}
\caption{ Imaginary part of the eigenfrequencies along the branch on which BIC appears;
(a) P-wave (lower branch) and (b) S-wave (upper branch).}
\end{figure}

\begin{figure}
\centering
\includegraphics[width=15cm]{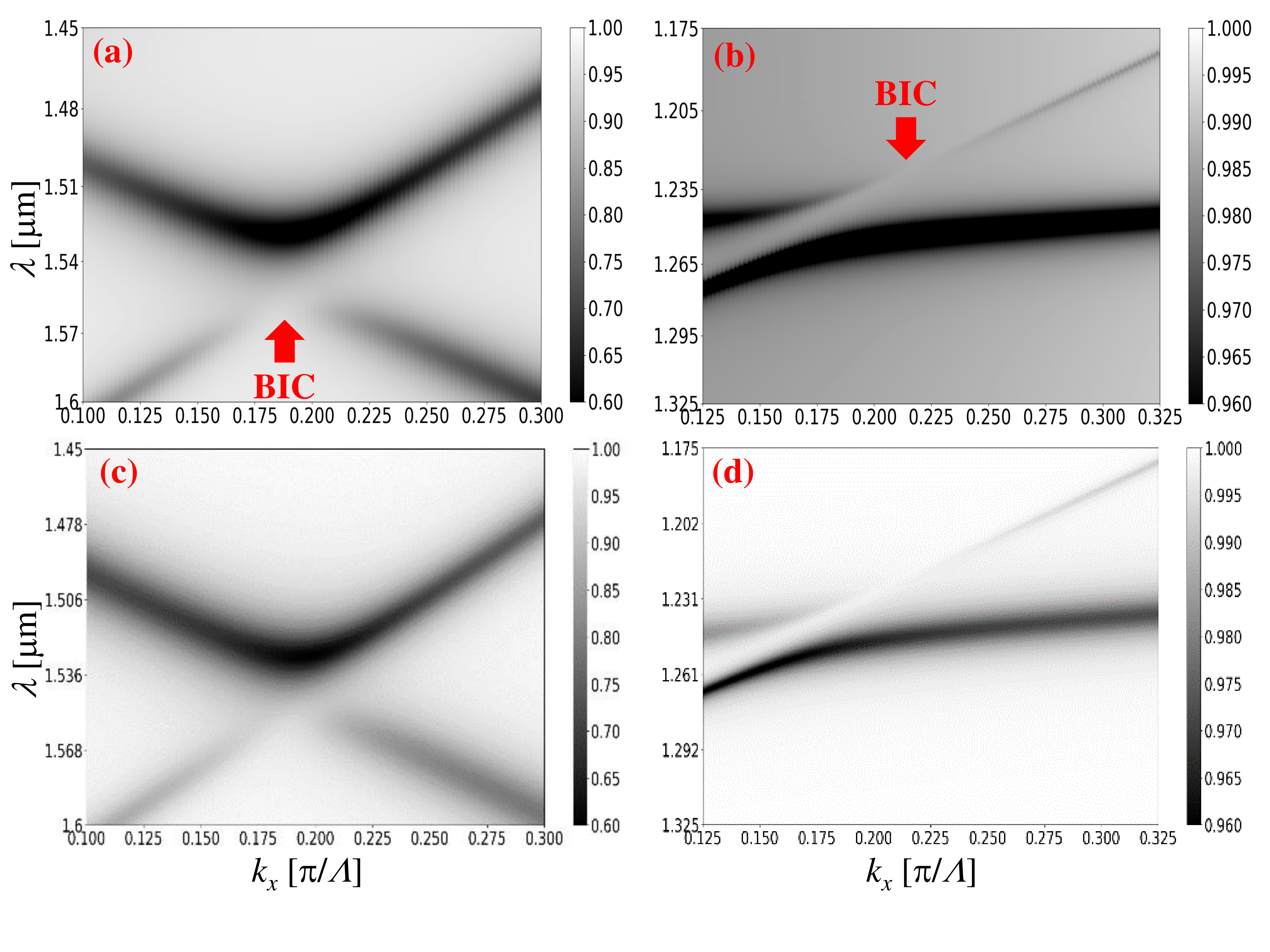}
\caption{Reflectance as a function of the in-plane wave vector and the incident wavelength  around the anti-crossing point. (a) P-wave and (b) S-wave redrawn from Figs. 2(a) and 2(b), respectively. The results of the TCMT calculation using Eq. (7) for (c) $p=1$ and (d) $p=-1$ with the following parameters; (c) $\gamma_{i1}/c=2.03\times10^{-2}, \gamma_{i2}/c=3.15\times10^{-2}, \gamma_{e1}/c=2.18\times10^{-3}, \gamma_{e2}/c=4.42\times10^{-3}, \alpha/c=3.52\times10^{-2}$, (d) $\gamma_{i1}/c=6.24\times10^{-3}, \gamma_{i2}/c=2.64\times10^{-2}, \gamma_{e1}/c=5.77\times10^{-5}, \gamma_{e2}/c=3.33\times10^{-4}, \alpha/c=3.28\times10^{-2}$. Here $c$ is the velocity of light in a vacuum.}
\end{figure}

\begin{figure}
\centering
\includegraphics[width=15cm]{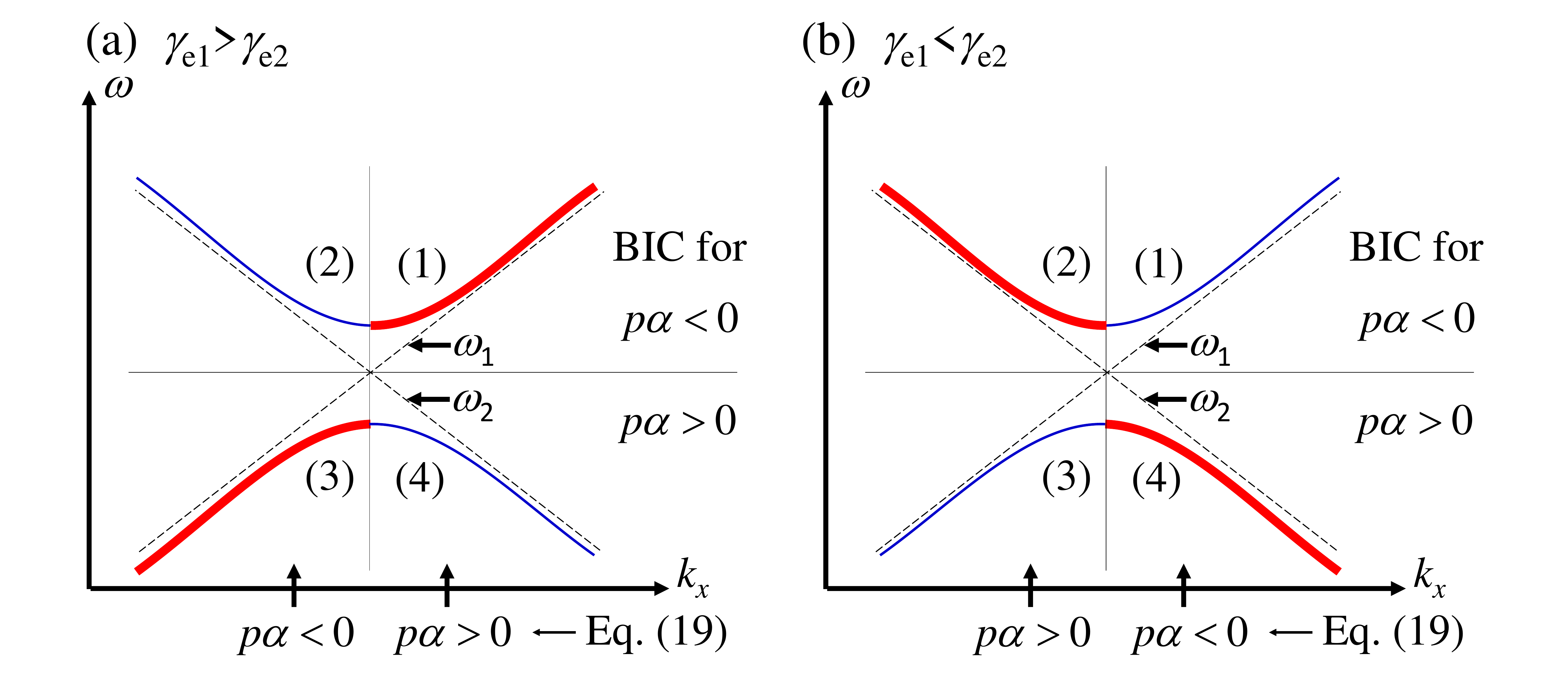}
\caption{Illustration  of the quadrant where BIC appears in the dispersion diagram near the anti-crossing point for (a)$\gamma_{e1}>\gamma_{e2}$ and (b)$\gamma_{e1}<\gamma_{e2}$.
The red and blue lines depict the dispersions whose radiation loss is the larger
(broad linewidth) and the smaller (narrow linewidth) of the two, respectively.}
\end{figure}

\begin{figure}
\centering
\includegraphics[width=15cm]{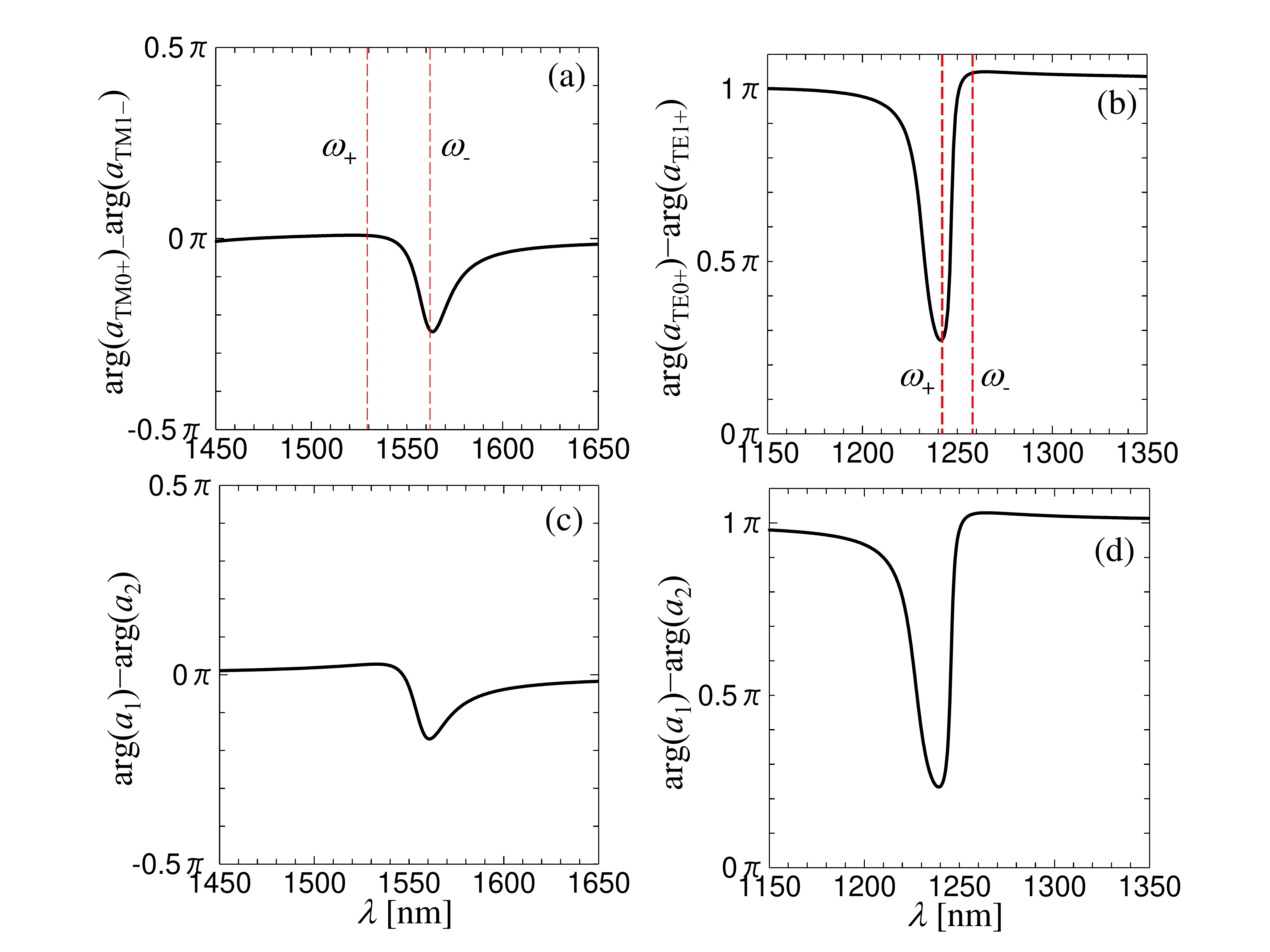}
\caption{Phase difference of two modes excited by the external radiation; (a) P-wave at  $k_{x}=0.1905$ $[\pi/\mathit{\Lambda}]$ and (b) S-wave at  $k_{x}=0.1600\ [\pi/\mathit{\Lambda}]$ by the SCMT calculations, (c)$p=1$ and (d)$p=-1$ by the TCMT calculations. Red dashed lines in (a) and (b) indicate the wavelengths of the coupled resonant modes $\omega_{+}$ and $\omega_{-}$.}
\end{figure}

\end{document}